\documentclass[twocolumn,amssymb, nobibnotes, aps, prc, superscriptaddress,floatfix, nobalancelastpage,longbibliography]{revtex4-2}

 \usepackage{graphicx}
\usepackage{bm}
\usepackage{amsmath} 
\usepackage{amssymb}
\usepackage{appendix}
\usepackage{braket} 
\usepackage{epsfig}
\usepackage{tensor}
\usepackage{CJKutf8}
\usepackage[version=4]{mhchem}
\usepackage{titlesec}
\usepackage{ifthen}
\usepackage{sidecap}
\usepackage{listings}
\usepackage{array}
\usepackage{floatrow}
\usepackage{enumitem}
\usepackage[para,online,flushleft]{threeparttablex}
\usepackage{orcidlink}
\usepackage[utf8]{inputenc}
\usepackage{float}
\floatstyle{plaintop}
\restylefloat{table}

\usepackage{hyperref}
\hypersetup{breaklinks=true,colorlinks=true,linkcolor=blue,citecolor=blue,filecolor=magenta,urlcolor=cyan}

\usepackage[all]{hypcap}

\usepackage{xcolor}
\definecolor{pastelgray}{rgb}{0.81, 0.81, 0.77}
\definecolor{beaublue}{rgb}{0.9, 0.9, 0.93}

\makeatletter
\def\@bibdataout@aps{%
\immediate\write\@bibdataout{%
@CONTROL{%
apsrev41Control%
\longbibliography@sw{%
    ,author="08",editor="1",pages="1",title="0",year="1"%
    }{%
    ,author="08",editor="1",pages="1",title="",year="1"%
    }%
  }%
}%
\if@filesw \immediate \write \@auxout {\string \citation {apsrev41Control}}\fi
}
\makeatother

\begin{document}

\title{Nucleonic shells and nuclear masses}


\author{Landon Buskirk}
\affiliation{Facility for Rare Isotope Beams, Michigan State University, East Lansing, Michigan 48824, USA}
\affiliation{Department of Physics and Astronomy, Michigan State University, East Lansing, Michigan 48824, USA}

\author{Kyle Godbey\,\orcidlink{0000-0003-0622-3646}}
\affiliation{Facility for Rare Isotope Beams, Michigan State University, East Lansing, Michigan 48824, USA}

\author{Witold Nazarewicz\,\orcidlink{0000-0002-8084-7425}}
\affiliation{Facility for Rare Isotope Beams, Michigan State University, East Lansing, Michigan 48824, USA}
\affiliation{Department of Physics and Astronomy, Michigan State University, East Lansing, Michigan 48824, USA}

\author{Wojciech Satu{\l}a\,\orcidlink{00000-0003-0203-3773}}
\affiliation{Institute of Theoretical Physics, Faculty of Physics, University of Warsaw, PL-02093 Warsaw, Poland}

\begin{abstract}
The binding energy of an isotope is a sensitive indicator of the underlying shell structure as it reflects the net energy content of a nucleus. Since magic nuclei are significantly lighter, or more bound, compared to their neighbors, the presence of nucleonic shell structure makes an imprint on nuclear masses. In this work, using a carefully designed  binding-energy indicator, we catalog the appearance of spherical and deformed shell and subshell closures throughout the nuclear landscape. After presenting experimental evidence for shell and subshell closures as seen through the lens of nuclear masses, we study the ability  of global nuclear mass models to predict local binding-energy variations related to shell effects. 
\end{abstract}

\date{\today}

\maketitle

\section{Introduction}

Nuclei with 2, 8, 20, 28, 50, 82, and 126 nucleons  have been found to be special  by having an exceptionally high natural abundance or being locally lighter than their neighbors \cite{Mayer1948}. These {\it magic} nucleon numbers were explained by the nuclear shell model  \cite{Mayer1949,Haxel1949} in terms of completely filled  nucleon shells. The nuclei with such numbers of nucleons are  referred to as magic, like  doubly-magic $^{48}_{20}$Ca$_{28}$  or semi-magic $^{120}_{~50}$Sn$_{70}$. 
Experimentally, there are numerous signatures of  magic gaps 
of shell closures. They include: locally enhanced binding energies, rapid changes of separation energies, 
low-lying collective excitations, kinks in charge radii, and spectroscopic factors, among other things \cite{Sorlin2008,Otsuka2020,Nowacki2021}.

The quantal stability of the atomic nucleus is  determined by the behavior of the single-particle level density $\rho(e)$ of the mean-field (intrinsic) Hamiltonian.
As the  ground state for many-fermion systems should  correspond to the lowest possible degeneracy, the nucleus is expected to be more bound if the  nucleonic level density near the Fermi level is low. Exceptionally stable systems (doubly magic nuclei) are indeed those with the least degenerate single-particle  level density around the Fermi level.
Quantitatively, the extra stability due to the presence of shell gaps can be encapsulated in the microscopic shell energy $E^{\rm shell}$ \cite{Strutinsky1968,Brack1972,Nil95aB} that
fluctuates with particle number  and reflects the non-uniformities of
the single-particle level distribution. Formally, the shell energy  can be approximated by:
\begin{equation}\label{EE}
E^{\rm shell} =\sum_{i=1}^Ae_i - \int e \tilde{\rho}(e)de,
\end{equation}
where $e_i$'s are  single-particle (Hartree-Fock) energies and $\tilde{\rho}(e)$ is the smoothed
single-particle density   that averages out single-particle energies within large
energy interval of the order of the energy difference between major shells.
The total binding energy
of a nucleus can be roughly given by \cite{Strutinsky1968,Brack1972}
\begin{equation}\label{LDP}
B= B^{\rm macr}+E^{\rm shell},
\end{equation}
where $B^{\rm macr}$ is the ``macroscopic'' energy that gradually depends on the number of
nucleons and thus associated with the smooth distribution of
single-particle  levels given by $\tilde{\rho}(e)$.

The behavior of $E^{\rm shell}$
changes periodically with  particle number. The lowest shell energy is expected
in the regions of low single-particle level density, e.g., 
for the spherical magic numbers 8, 20, 28, 50, 82, and 126.
However,  below and above these magic    numbers, the  level density becomes large
[(2$j$+1)-fold   degeneracy of spherical orbitals] and a Jahn-Teller transition takes place  towards
deformed shapes \cite{Reinhard1984,Nazarewicz1994}. The stabilisation of deformed nuclei can be associated with 
energy gaps  in deformed   s.p. levels, i.e., deformed sub-shell closures \cite{Brack1972,Ragnarsson1984,Nil95aB}.
Examples of deformed s.p. diagrams, or Nilsson plots, can be found in, e.g.,
 Appendix on Nuclear Structure of 
Ref.~\cite{Firestone1999}.

\section{Binding-energy indicators}

 Empirical information on the magnitude of  nucleonic correlations
 is often extracted from experimental data using
binding-energy  relations
 (filters, indicators) based on measured masses of neighboring nuclei \cite{JANECKE1985,JENSEN1984}.
 

Usually, the binding-energy indicators are the finite-difference
expressions representing various derivatives of (positive) nuclear binding energy $B(N,Z)$
with respect to $N$ and $Z$.  Their role is to isolate some
specific parts of the correlation  energy by filtering out  that part of
the binding energy which behaves as a polynomial of a required
order in $N$ and $Z$.  
The commonly used mass differences are 
one-nucleon separation energies $S_\tau$ ($\tau=n,p$).
For neutrons:  
\begin{equation}\label{Sn}
 	S_n (N, Z) = B(N, Z) - B(N-1, Z).
\end{equation} 
The  two-neutron separation energy is
\begin{equation}\label{S2n}
 	S_{2n} (N, Z) = B(N, Z) - B(N-2, Z).
\end{equation}
The difference $\delta_{2n}=S_{2n}(N,Z) - S_{2n}(N+2,Z)$  is
the so-called two-neutron shell gap indicator that represents twice the gap in the corresponding single-particle spectrum \cite{Bender2002}. 
The neutron chemical potential $\lambda_n$ can be expressed through two-neutron separation energies
\cite{Beiner1974,Beiner1975,Bengtsson1984}:
\begin{equation}\label{lambda}
\lambda_n(N-1,Z)  \approx   -\frac{1}{2}  	S_{2n} (N=2k, Z),
\end{equation}
where $2k$ indicates an even number. We note that $\lambda_n$ is  negative for bound systems.
In addition, 
\begin{align}\label{lambdadelta}
S_{n} (N=2k, Z) \approx &  -\lambda_n(N-1,Z)  -\frac{1}{2}  \frac{\partial\lambda_n(N-1,Z)}{\partial N }\nonumber \\
 &+ \Delta_n(N-1,Z),
\end{align}
where $\Delta_n(N-1,Z)$ is the average neutron pairing gap \cite{Beiner1974,Beiner1975}. 

The single-particle (s.p.) neutron energy splitting at the Fermi level, $\Delta  e_n$, 
can thus be defined in terms of one-nucleon separation energy differences
\cite{Satula1998,Dobaczewski2001}:
\begin{equation}\label{eq:DEn}
    \Delta  e_n(N=2k,Z)= S_n(N,Z) - S_n(N+2,Z).
    \end{equation}
As demonstrated in Refs.~\cite{Satula1998,Dobaczewski2001}, if variations of the mean field and pairing are smooth along isotopic or isotonic chains, the filter $\Delta  e_\tau$  represents the energy difference between the lowest particle level and the highest hole (occupied) level. For instance:
\begin{equation}\label{esp}
\Delta  e_n(N=2k,Z) \approx e^n_{k+1}-e^n_{k}. 
    \end{equation}
Similar relations to Eqs.~(\ref{Sn} - \ref{esp}) hold for protons.
It directly follows   from Eqs.~(\ref{lambdadelta}) and (\ref{eq:DEn}) that
$\Delta  e_\tau$ is proportional to the derivative of $\lambda_\tau$ with respect to the particle number $N_\tau$ ($N_\tau=Z$ or $N$ for $\tau=p$ or $n$), i.e., it is inversely proportional to the level density \cite{Bengtsson1981}.
 The indicator $\Delta{e}_\tau$ is  thus sensitive to small changes of the level density at the Fermi level. Indeed, the regions of the low level density 
are expected to correspond to  increased values of $\Delta  {e}_\tau$. 

Since for the smoothly varying mean-field potentials the chemical potential gradually {\it increases}
with particle number, $\Delta  e_\tau$  should be positive in general.
The deviations from the monotonic behavior of $\lambda_\tau(N_\tau)$ do occur, and  are usually associated with the rapid change of nuclear mean fields due to configuration changes. In some cases, usually associated with shape transitions, $\Delta  e_\tau <0$; this corresponds to a backbending in the gauge space of $N_\tau(\lambda_\tau)$ \cite{Bengtsson1981,Bengtsson1984,Zhang1984}.

\begin{figure}[!htbp]
    \centering{\includegraphics[width=1.0\textwidth]{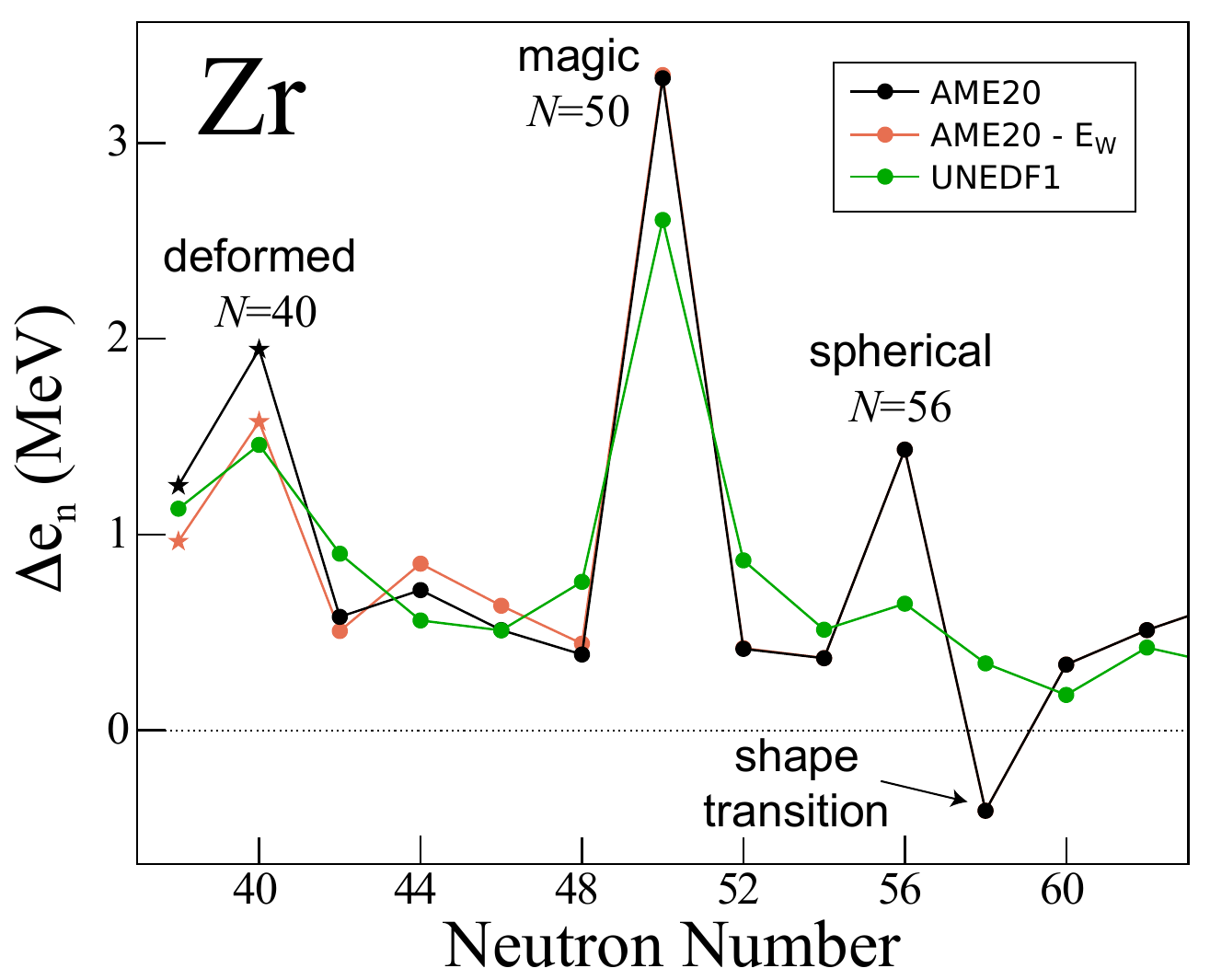}}
    \caption{Experimental values and model predictions of $\Delta  {e}_n$ across zirconium isotopes. Extrapolated values from experimental data are marked with stars. Strong peaks appear for the deformed gap at N=40, the magic gap at N=50, and the spherical gap at N=56. See text for details.}
    \label{fig:ZrChain}
\end{figure}

As an illustrative example, Fig.\,\ref{fig:ZrChain} shows $\Delta  \tilde{e}_n$ for the Zr  isotopic chain.
The local maxima in $\Delta  {e}_n$ can be associated with spherical and deformed s.p. gaps
discussed in Sec.~\ref{sec:trends}. The negative value of $\Delta  {e}_n$ at $N=58$ reflects the well-known spherical-to-deformed shape transition around $^{98}$Zr
\cite{Heyde2011,Reinhard1999}.

While the goal of our work is to demonstrate  that $\Delta  e_\tau$ is a superb measure of 
spherical and deformed shell closures, this indicator can also be used  to study mean level spacing, or  mean level density, at the Fermi energy. Indeed, beyond the regions of low level density associated with gaps,
$\Delta  e_\tau$ represents mean level splitting
at the Fermi energy. In the simplest scenario assuming Kramers and isospin degeneracy, the mean level spacing equals $\bar \varepsilon = 4/{\tilde\rho}(\lambda)$, where 
${\tilde\rho}(\lambda) = 6a/ \pi^2$ and $a$ stands for the level density parameter, the value of which is uncertain.  In the simplest isoscalar scenario assuming dominant volume-like 
$A$-dependence the estimates for $a$ vary from $A/10$ (which is the harmonic oscillator limit \cite{(Boh69)}) to  $A/8$\, MeV$^{-1}$~\cite{(Gil65),(Kat80),(Shl92)}. This, in turn, gives $\bar \varepsilon \approx (60\pm 6)/A$\, MeV. Note, that for the Zr isotopic chain 
presented in Fig.\,\ref{fig:ZrChain} it varies from  0.75(8)\, MeV for $A=80$ to
0.60(6)\, MeV for $A=100$. The estimates agree relatively well with the data 
shown in Fig.\,\ref{fig:ZrChain} outside the regions of low level density associated with deformed and spherical energy gaps.

\section{Datasets and models}

In our analysis we use the most recent measured values of nuclear binding energies from the AME2020 dataset \cite{AME2020}. 
In this analysis we do not consider experimental errors and theoretical uncertainties as their proper inclusion would require the knowledge of underlying covariances.
While in many nuclei the experimental mass errors are well below theoretical uncertainties and can be ignored, this is no longer the case for very exotic nuclei far from stability.
In the simplest case where one assumes completely uncorrelated errors, the total error of mass filters grows substantially as several masses are involved.
A detailed error analysis of $\Delta  {e}_\tau$ and other mass filters will be a subject of forthcoming study.

As for prediction, 
we consider seven theoretical models based on the energy density functional theory (EDF) which are capable of   describing  
the whole nuclear chart: SkM$^*$ \cite{Bartel1982}, 
SkP \cite{Dob84}, 
SLy4 \cite{Chabanat1995}, 
SV-min \cite{Kluepfel2009}, 
UNEDF0 \cite{UNEDF0}, 
UNEDF1 \cite{UNEDF1}, and
UNEDF2 \cite{UNEDF2}. 
The above set of EDF models was augmented by a  well-fitted  mass  model FRDM-2012 \cite{Moller2012}
that  has significantly more parameters than the (less phenomenological) DFT models, 
resulting in a
better fit to measured masses.

For $\Delta  e_\tau$  extraction from the data, the Wigner energy has to be removed from experimental binding energies
as it represents an irregularity (kink) in the mass surface around $|N=Z|$ and hence impacts mass difference indicators aiming at extracting shell structure effects \cite{Gelberg2009,Hamaker2021}.
In Ref.~\cite{SATULA1997}, the Wigner term has been parameterized as 
\begin{equation}\label{eq:Ew2}
    E_W(2) = a_W|N-Z|/A,
\end{equation}
where $a_W = 47$\,MeV. However, this expression notably underestimates the Wigner energy for $^{80}$Zr and $^{56}$Ni, two locations of shell closures that are later discussed. For this reason, we supplement $E_W(2)$ with the model of Ref. \cite{Goriely2013}:
\begin{equation}\label{eq:Ew1}
    E_W(1) = V_We^{-\lambda_W(\frac{N-Z}{A})^2} + V'_W|N-Z|e^{-\left(\frac{A}{A_0}\right)^2}
\end{equation}
where $V_W = 1.8$\,MeV, $\lambda_W$ = 380, $V'_W =-0.84$\,MeV, and $A_0= 26$. In our analysis, the average of $E_W(1)$ and $E_W(2)$ has been subtracted from all experimental binding energies. The effect of such subtraction is illustrated in Fig.\,\ref{fig:ZrChain} for $\Delta  {e}_n$ along
the Zr chain (see Ref.~\cite{Hamaker2021} for the discussion of the $^{80}$Zr case).

\section{BMEX web application}

The exploration of the experimental and theoretical data was performed using the Bayesian Mass Explorer (BMEX)~\cite{bmex} web application and the associated database.
An evolution of the Mass Explorer project~\cite{massexplorer}, BMEX contains a suite of online plotting and comparison tools that were used to produce the draft figures in the current work.
The BMEX database and software are hosted in a cloud computing environment and do not require any downloads or installation by the end user to access the tool.
To save the user's sessions, plot exporting and link sharing is also included without the need for any user accounts or logins. A screenshot of the application can be found  in Fig.~\ref{fig:bmex} in Appendix~\ref{sec:bmex}.


\begin{figure*}[htbp]
    \centering{\includegraphics[width=0.8\textwidth]{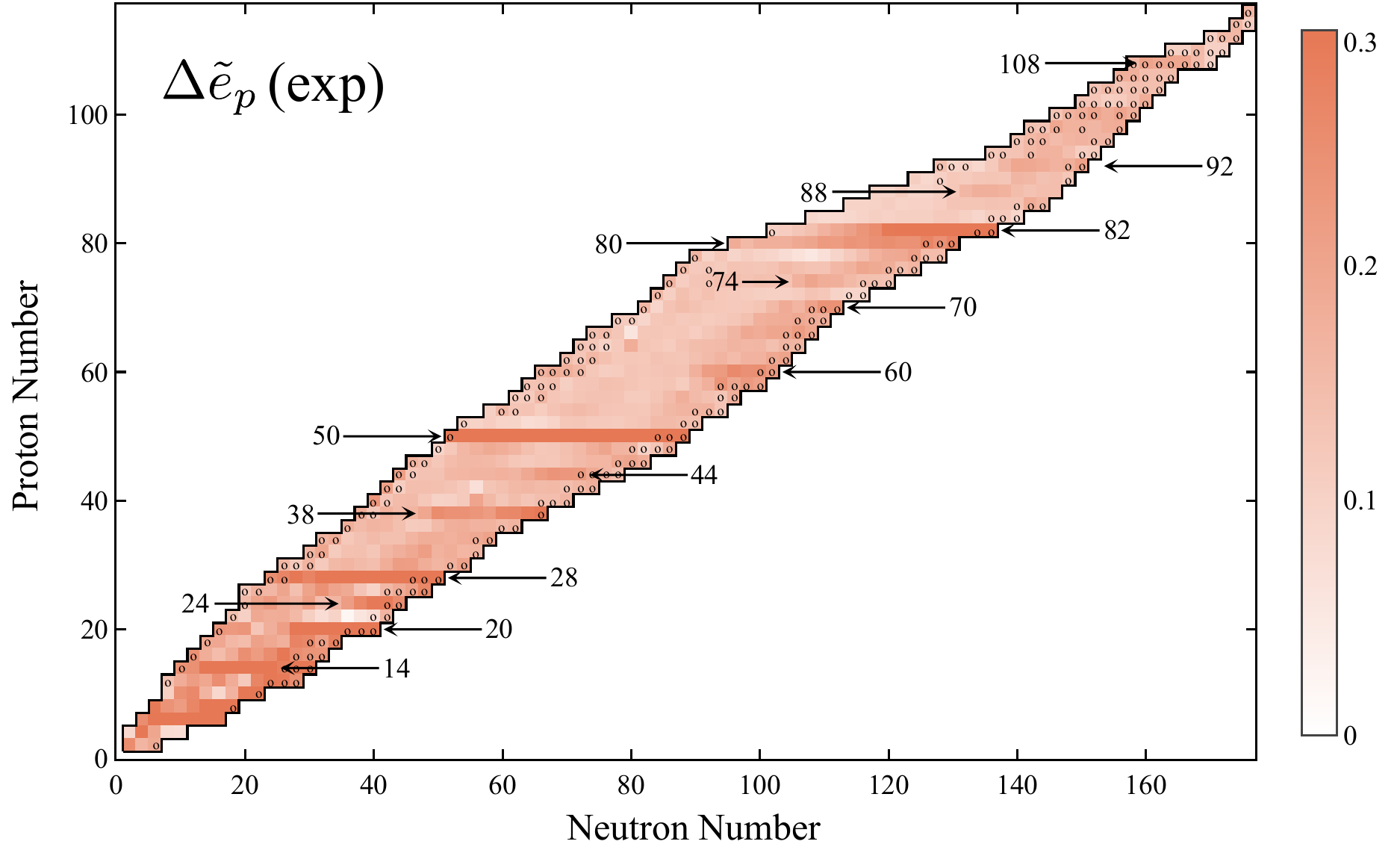}}
    \caption{Experimental values of the dimensionless splitting $\Delta  \tilde{e}_p$ 
    (\ref{enorm}) throughout the nuclear landscape. The nuclei for which the expression (\ref{esp}) involves binding energies extrapolated from systematic trends in \cite{AME2020} are marked by circles.  Shell closures corresponding to the bands of locally elevated values of $\Delta  \tilde{e}_p$ are clearly seen.}
    \label{fig:AME2020_SPES}
\end{figure*}

\section{Systematic trends}\label{sec:trends}

In order to remove the average mass and isospin dependence of shell gaps, we 
scale
 $\Delta  e_\tau$ by
the average oscillator frequency\,\cite{Nilsson1969}: 
\begin{equation}\label{eq:hw0}
    \hbar\omega_{0} =  41A^{-1/3} (1 \pm \frac{N-Z}{3A})~{\rm MeV},
\end{equation}
where the plus sign holds for neutrons and the minus sign for protons. In the following, we discuss the dimensionless splittings 
\begin{equation}\label{enorm}
\Delta  \tilde{e}_\tau \equiv \Delta  e_\tau/\hbar\omega_{0}.
\end{equation}
When interpreting the patterns of shell gaps in the $(N,Z)$ plane, it is important to  recall that nuclei 
close to the spherical magic gaps at $Z=20$, 28, 50, 82, and 126 are nearly spherical and that the quadrupole collectivity primarily depends on the distance of $Z$ and $N$ to the closest magic proton and neutron number \cite{Janecke1981,Pritychenko2016}. That is, the largest quadrupole deformations are expected in the regions between spherical magic gaps.

\begin{figure*}[htbp]
    \centering{\includegraphics[width=0.8\textwidth]{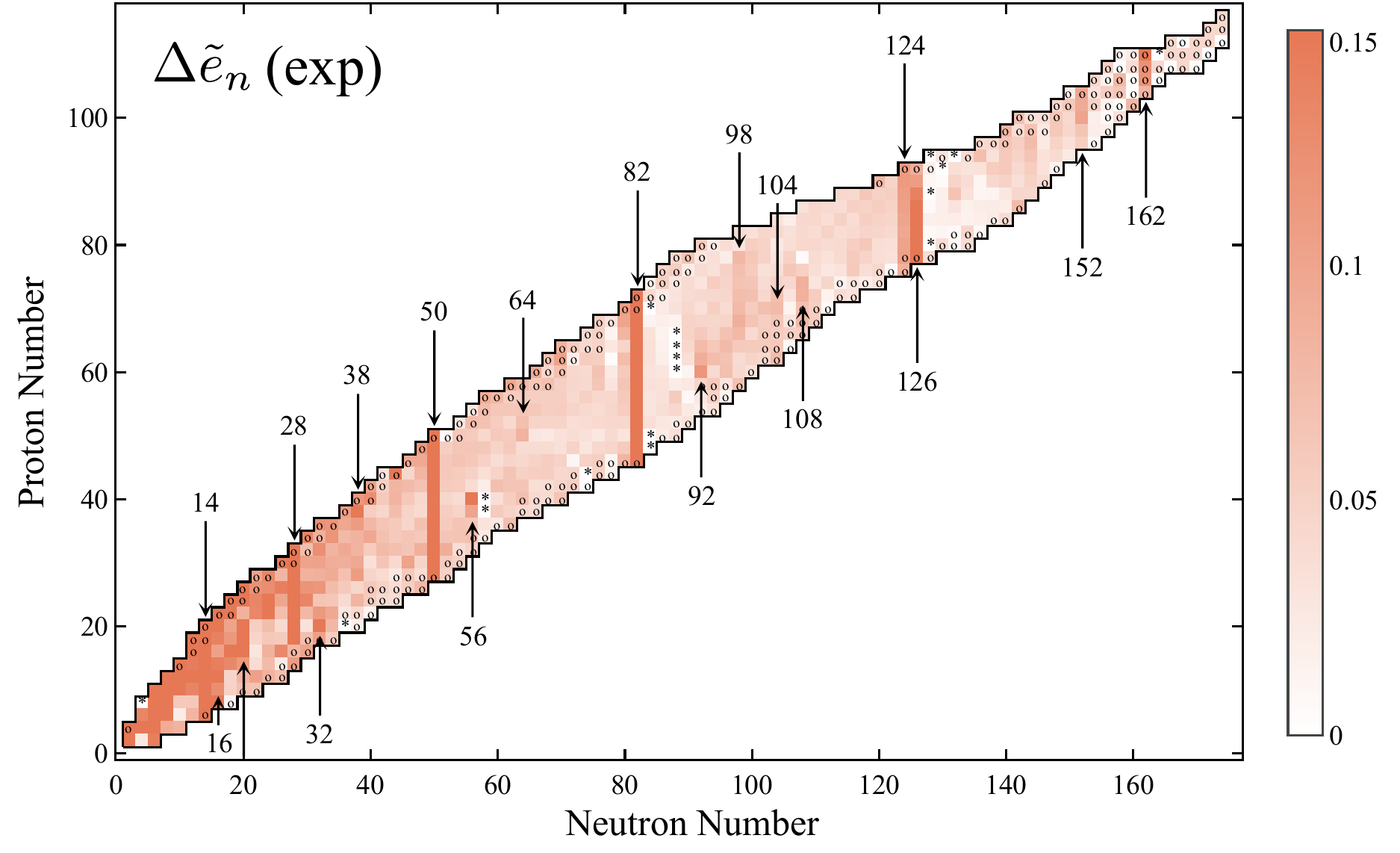}}
    \caption{Similar as in Fig.~\ref{fig:AME2020_SPES} but for $\Delta  \tilde{e}_n$.
    The nuclei with negative values of $\Delta  \tilde{e}_n$ are marked by an asterisk.}
    \label{fig:AME2020_SNES}
\end{figure*}

\subsection{Experimental single-nucleon shell gaps}
Figure~\ref{fig:AME2020_SPES} shows the proton shell gaps $\Delta  \tilde{e}_p$ extracted from experimental binding energies. The experimental neutron shell gaps $\Delta  \tilde{e}_n$ are displayed in Fig.\,\ref{fig:AME2020_SNES}. 
The spherical magic  gaps are clearly seen for both protons and neutrons. In addition, isotopic and isotonic bands 
of locally enhanced values of $\Delta  \tilde{e}_\tau$ are present; they
can be associated with local subshell closures, both spherical and deformed. 
They are discussed in the following.

\subsubsection{Spherical magic gaps} 
In the protons, the pronounced $Z=50$ gap extends across the nuclear landscape. The $Z=82$ gap is large for $N\ge 126$ but it seems to gradually fade away in neutron deficient Pb isotopes. This is consistent with the presence of shape coexistence effects in these nuclei, in which spherical, prolate, and oblate structures coexist (and interact) at low energies \cite{Wood1992,Heyde2011}. While the  $Z=28$ proton shell gap is generally pronounced, the $Z=20$ gap becomes fairly diluted below $N=24$.

The neutron magic gaps $N=50, 82$, and 126 are well pronounced. The $N=28$ gap
deteriorates in the lightest isotones, and a similar situation is seen at $N=20$. The disappearance of $N=20$ and 28 magic gaps in neutron-rich nuclei is supported by
an  appreciable experimental evidence for deformed structures  below $^{44}$S  and  $^{32}$Mg \cite{Sorlin2008,Heyde2011}.

\subsubsection{Spherical subshell closures} 
Several local spherical shell gaps can be identified in
Figs.\,\ref{fig:AME2020_SPES} and \ref{fig:AME2020_SNES}.
They include:
 $Z=14$ subshell closure  in the Si isotopes \cite{Cottle2007};
$Z=64$ subshell closure  in $^{146}$Gd \cite{Bengtsson1984};
$N=16$ subshell closure in $^{36}$Ca \cite{Lalanne2023}  and $^{24}$O \cite{Tshoo2012};
$N=32$ subshell closure in $^{52}$Ca \cite{Wienholtz2013};
$N=56$ subshell closure in $^{96}$Zr \cite{Molnar1991}; and
$N=64$ subshell closure in Sn \cite{Piller1990}.
The single $2p_{1/2}$  orbital separates the $N=126$ magic gap from the $N=124$ spherical subshell \cite{Vergnes1990}. Consequently,  these two shell closures overlap in Fig.\,\ref{fig:AME2020_SNES}.

\subsubsection{Deformed subshell closures} 
In the regions between spherical magic gaps, 
the indicator $\Delta  \tilde{e}_\tau$  provides important information about deformed shell gaps.
The region of deformed nuclei around $^{64}$Cr \cite{Sorlin2003,Babcock2016,Silwal2022} 
can be associated with the deformed subshell closures  $Z=24$ and $N=40$ \cite{Oba2008}. In Fig.\,\ref{fig:AME2020_SPES}, the proton shell gap $\Delta  \tilde{e}_p$ is well pronounced for neutron-rich Cr isotopes.
Of particular interest are deformed shell closures at  $Z=38, 40$ that are responsible for very large ground-state deformations around $^{76}$Sr \cite{Nazarewicz1985}, $^{80}$Zr \cite{Hamaker2021}, and
$^{102}$Zr \cite{Heyde2011}.
The $Z=80$ oblate gap is responsible for weakly deformed ground states of the Hg isotopes \cite{Vergnes1990}. It is separated from the $Z=82$ magic gap by a single $2s_{1/2}$ orbit so these two shell closures overlap in Fig.\,\ref{fig:AME2020_SPES}.

The deformed neutron gaps in the rare-earth nuclei seen in  Fig.\,\ref{fig:AME2020_SNES} include:
$N=98$ gap in the Gd-Dy region \cite{Hartley2018,Bengtsson1981};
$N=104$ gap around $^{174}$Yb  \cite{Bengtsson1984}; and
$N=108$ gap known around $^{182}$W \cite{Bengtsson1981}.

In the actinide and transfermium regions, the most pronounced deformed neutron closures are $N=152$ 
\cite{Ramirez2012,Makii2007} and $N=162$ \cite{Bengtsson1984,Dvorak2006,Oganessian2013,Kaleja2022}. 
In the protons, the deformed shell gap at $Z=108$ is particularly pronounced
\cite{Patyk1991,Moller1992,Cwiok1994,Dvorak2006,Oganessian2013}.
These subshells are essential for the stabilization of nuclear binding in the transactinides.

In addition to the above list of shell and subshell closures that can be straightforwardly identified, there are  other regions in Figs.\,\ref{fig:AME2020_SPES} and \ref{fig:AME2020_SNES} with moderately enhanced values of $\Delta  \tilde{e}_\tau$. For instance,
the $N=92$ shell effect around $^{152}$Nd can probably be attributed to octupole correlations.

\subsubsection{Shape transitions}
 Negative values of $\Delta  {e}_\tau$ are associated with shape transition.
Several regions of shape-transitional behavior are seen in Fig.\,\ref{fig:AME2020_SNES}.
They include the region of shape coexistence around $^{98}$Zr and the 
transition regions to well deformed prolate shapes around $N=88$ \cite{Bengtsson1981,Heyde2011}.

It is interesting to notice that rapid shape transitions are clearly seen in $\Delta  \tilde{e}_n$ 
 in Fig.\,\ref{fig:AME2020_SNES}
but not in $\Delta  \tilde{e}_p$. Indeed,  no regions of $\Delta  \tilde{e}_p<0$ can be seen in Fig.~\ref{fig:AME2020_SPES}, which indicates that the proton chemical potential $\lambda_p$ increases
monotonically with $Z$ throughout the nuclear landscape.

\subsubsection{Two-nucleon shell gap indicator}
The plots of experimental $\delta_{2\tau}$  are shown in Figs.~\ref{fig:SPESd} and \ref{fig:SNESd} in Appendix~\ref{sec:d2tau}.  As seen, this indicator   behaves in a similar way as $\Delta  {e}_\tau$, though,  in practice, the resolving power of the $\delta_{2\tau}$  for identifying subshell closures is slightly below that of  $\Delta  {e}_\tau$. Indeed, as shown in Ref. \cite{Gelberg2009},
$\delta_{2\tau}=\Delta  {e}_\tau + \Delta P_\tau $ where
$\Delta P_\tau $ represents a pairing correction. 
Consequently, $\delta_{2\tau}$ is more affected by  correlations, which tend to smear out shell effects.

\begin{figure*}[!htbp]
    \centering{\includegraphics[width=0.8\textwidth]{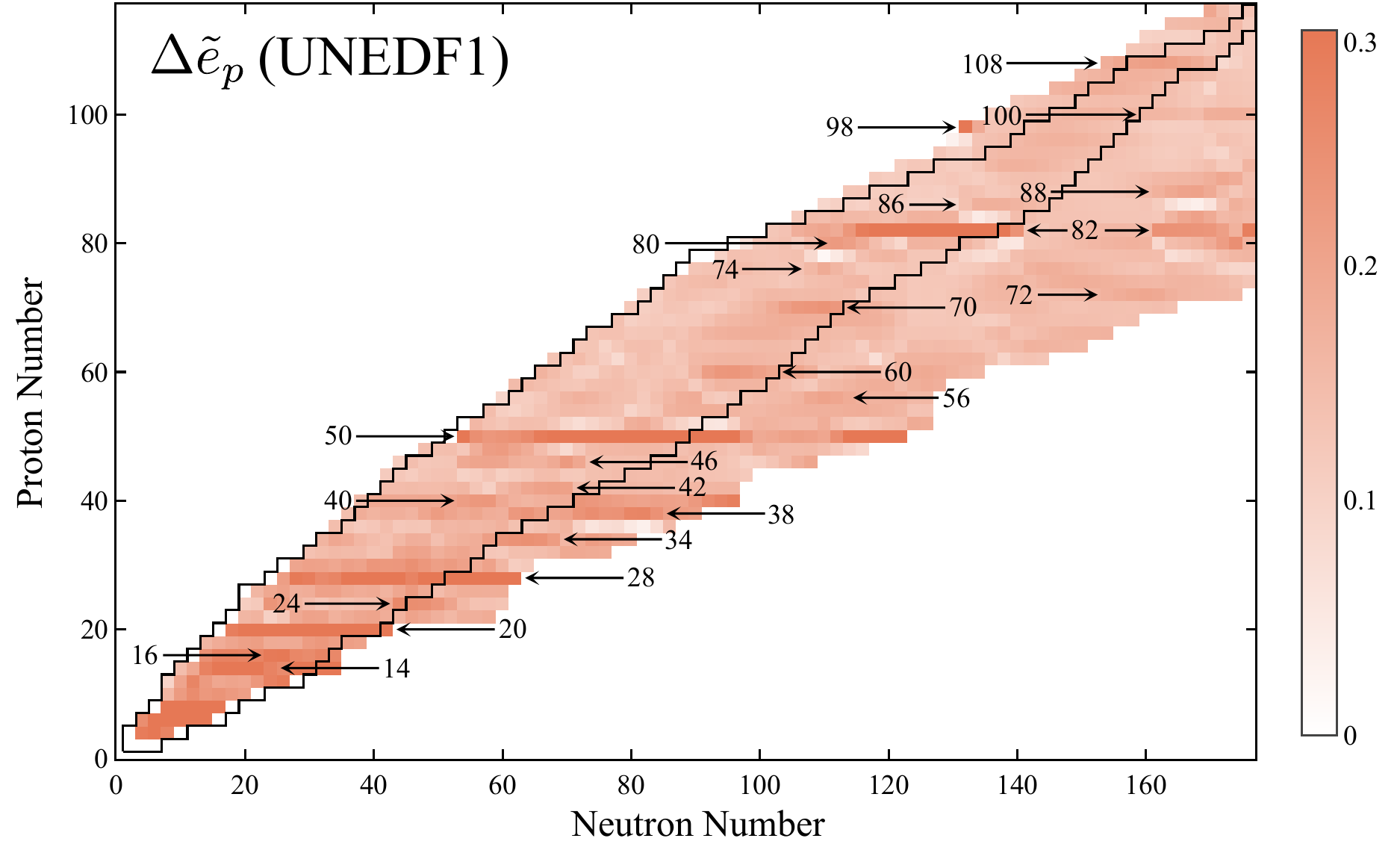}}
    \caption{Similar as in Fig.~\ref{fig:AME2020_SPES} but for the mass model  UNEDF1. The range of experimental data is marked by a solid black line.}
    \label{fig:UNEDF1_SPES}
\end{figure*}

\subsection{Model predictions}

 \begin{figure*}[htb]
    \centering{\includegraphics[width=0.8\textwidth]{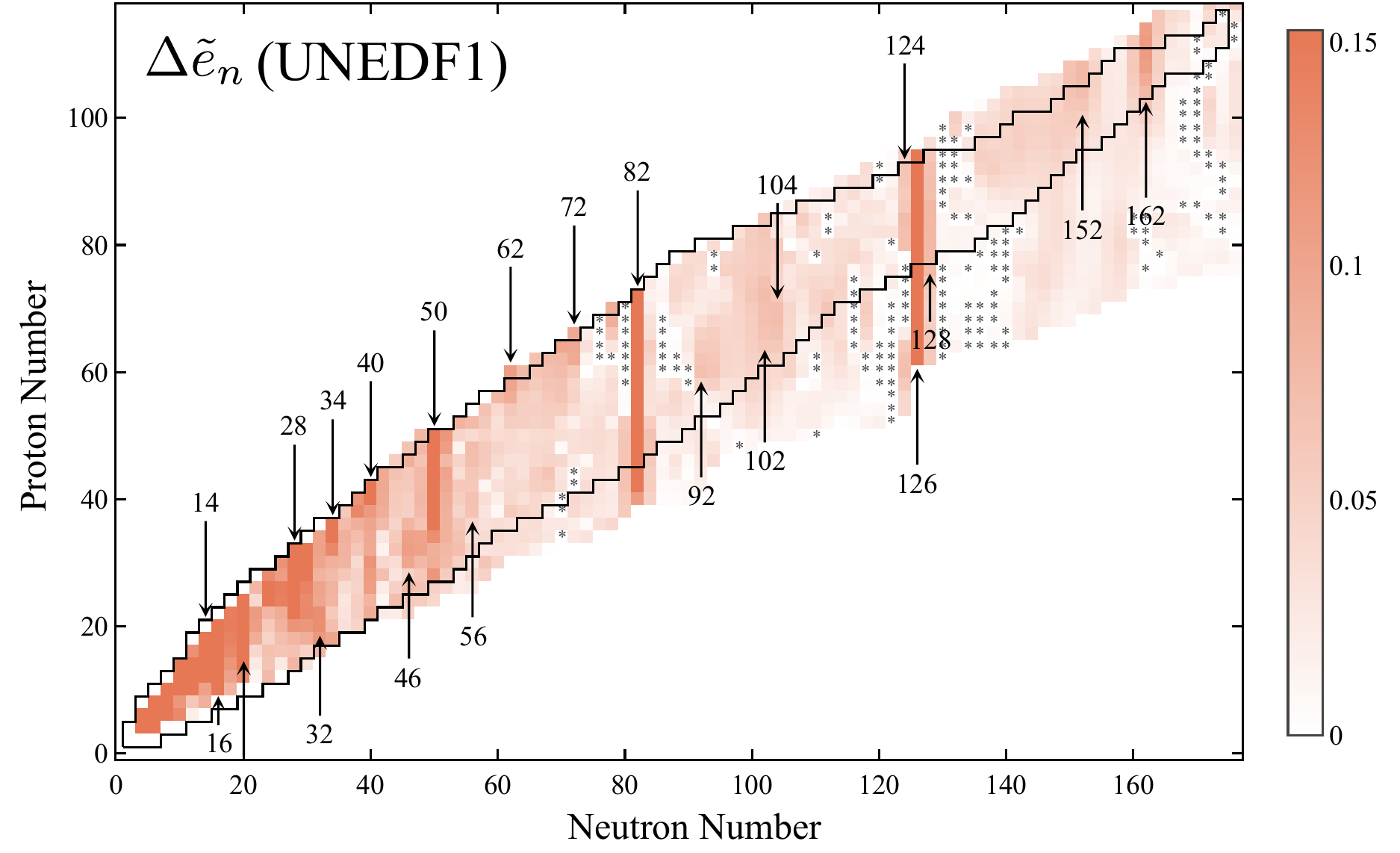}}
    \caption{Similar as in Fig.~\ref{fig:AME2020_SNES} but for the mass model  UNEDF1. The range of experimental data is marked by a solid black line and asterisks denote negative values.}
    \label{fig:UNEDF1_SNES}
\end{figure*}

Figures~\ref{fig:UNEDF1_SPES} and~\ref{fig:UNEDF1_SNES} illustrate the performance of the representative UNEDF1 mass model with respect to $\Delta \tilde{e}_p$ and $\Delta \tilde{e}_n$ respectively.
The predictions of other models can be obtained by using the BMEX tool \cite{bmex}.
The predictions extend beyond the region of nuclei with experimentally-known masses, and hence provide useful guidance for the future experiments at radioactive ion beam facilities. For instance, it is seen that the magic gaps $Z=50$ and $Z=82$   are significantly weakened around $N=106$ and $N=150$, respectively. 

\begin{table}[htb!]
    \caption{ \label{Performance}Performance of different mass models with respect to $\Delta \tilde{e}_\tau$
    corresponding different subshell closures seen in experimental data.
    The models are: SM=SkM$^*$, SP=SkP, SL=SLy4, SV=SV-min, U0=UNEDF0, U1=UNEDF1, U2=UNEDF2, and FR=FRDM-2012.}
\begin{ruledtabular}
    \begin{tabular}{ccccccccc}
        $\Delta {e}_\tau$ & SM & SP & SL &SV & U0 &U1 &U2 &FR\\
        \hline \\[-6pt]
        \multicolumn{9}{c}{Protons}\\[3pt]  {38}&\checkmark&&\checkmark&\checkmark&&\checkmark&\checkmark&\checkmark\\{40}&\checkmark&\checkmark&\checkmark&\checkmark&&\checkmark&\checkmark&\\
        {44}&&&\checkmark&&&&&\checkmark\\   {70}&\checkmark&\checkmark&\checkmark&\checkmark&\checkmark&\checkmark&\checkmark&\checkmark\\
        {74}&&\checkmark&&&&&&\checkmark\\{80}&\checkmark&&\checkmark&\checkmark&\checkmark&\checkmark&\checkmark&\checkmark\\
        {88}&&\checkmark&&&&&&\\{92}&\checkmark&\checkmark&\checkmark&\checkmark&\checkmark&\checkmark&\checkmark&\checkmark\\{108}&\checkmark&\checkmark&\checkmark&\checkmark&\checkmark&\checkmark&\checkmark&\checkmark\\
        \hline\\[-6pt]
        \multicolumn{9}{c}{Neutrons}\\[3pt]{56}&\checkmark&\checkmark&&\checkmark&\checkmark&\checkmark&\checkmark&\checkmark\\{70}&\checkmark&\checkmark&\checkmark&\checkmark&\checkmark&\checkmark&\checkmark&\checkmark\\{92}&\checkmark&&\checkmark&\checkmark&\checkmark&\checkmark&\checkmark&\checkmark\\
        {98}&&&&&&&&\checkmark\\{104}&\checkmark&\checkmark&\checkmark&\checkmark&\checkmark&\checkmark&\checkmark&\\
        {108}&&&&&&&&\checkmark\\{124}&&\checkmark&&\checkmark&\checkmark&\checkmark&\checkmark&\checkmark\\{152}&\checkmark&&\checkmark&&&\checkmark&\checkmark&\checkmark\\{162}&\checkmark&\checkmark&\checkmark&\checkmark&\checkmark&\checkmark&\checkmark&\checkmark\\
    \end{tabular}
    \end{ruledtabular}
\end{table}

The overall performance of the mass models with respect to $\Delta \tilde{e}_\tau$ is illustrated in Table~\ref{Performance}. As expected, FRDM-2012 performs fairly well overall. Several deformed subshell closures are robustly predicted in almost all models: $Z=70, 80, 92, 108$ and $N=92, 104$, and 162.
The same holds for spherical subshell closure $N=56$. While the peak at $N=56$ predicted by UNEDF1 is underestimated in Fig.~\ref{fig:ZrChain} the subshell closure is clearly seen in Fig. \ref{fig:UNEDF1_SNES}.

Other shells are predicted by a subset of models. In some cases, the ``theoretically-fragile'' gaps have been discussed  discussed in literature. See, e.g., Ref. \cite{Dobaczewski2015} for the $N=        152$ gap predictions.
Interestingly, the models consistently predict deformed proton shell gaps at $Z=46$ around $N=70$ and $Z=56$ around $N=72$, and the deformed neutron gap $N=72$ around $ Z=62$. These features are not clearly seen in the  experimental data.
In general, the predictive power of the mass models used in this study with respect to $\Delta \tilde{e}_\tau$
is quite reasonable. Moreover, the experimental finding that  $\Delta \tilde{e}_p$ is usually positive is nicely confirmed by theory, see Fig.~\ref{fig:UNEDF1_SPES}. The predicted regions of $\Delta \tilde{e}_n<0$ in Fig.~\ref{fig:UNEDF1_SNES} are broader than in experiment. This is to be expected as the shape transitions predicted by mean-field models are too abrupt  due to the missing dynamical (zero-point) correlations.
While the mean-field models are generally expected to reproduce shell and subshell closures at correct particle numbers, the actual size of the predicted $\Delta \tilde{e}_\tau$ is expected to depend on zero-point correlations and small model differences (e.g., due to poorly known spin-isospin terms \cite{Reinhard1999,Dobaczewski2015} of EDFs).


\section{Summary}

The s.p. energy splitting at the Fermi level $\Delta  e_\tau$ has been extracted from measured nuclear masses and compared with predictions of mean-field models. As  demonstrated  in this work,  $\Delta  e_\tau$ is indeed  a superb indicator of shell closures in spherical and deformed nuclei. In particular, this quantity can be very useful when studying the appearance and disappearance of nucleonic shell gaps in exotic nuclei. 

After cataloging experimental shell and subshell closures obtained by means of $\Delta \tilde{e}_\tau$, we showed  
that  EDF-based  models yield the placements of s.p. energy splitting maxima consistently with experiment. Indeed, mean-field models are expected to
perform well in this regard as the concept of intrinsic s.p. orbits and energies  is naturally present there. In some cases, such as the deformed $A\approx 80$
and $A\approx 100$ regions, theory sometimes poorly predicts the spherical-to-deformed shape transition due to missing zero-point correlations \cite{Reinhard1999}. This deficiency of current models
will need to be addressed. In this context, we wish to emphasize that  the intent of this work in not to exhaustively quantify the fidelity of theoretical models' predictions of $\Delta {e}_\tau$. 

Additionally, this work highlights the potential for user-focused scientific software to aid discovery and provide guidance for future experimental campaigns.
To this end, the BMEX tool used in this work  will be continually updated to include new experimental data and extended to a broader set of nuclear models.
A broader set of uncertainty estimates for both experimental and theoretical data will also be added to the tool. The new features will include estimates of experimental and theoretical errors on mass filters, and a Bayesian model mixing module that will combine the knowledge from multiple models \cite{neufcourt2019neutron,BAND,Kejzlar2023}.

\section*{Acknowledgements}

This material is based upon work supported by the U.S.
Department of Energy, Office of Science, Office of Nuclear
Physics under Awards Nos. DE-SC0023688  and DOE-DE-SC0013365,
the National Science Foundation under award number 2004601 (CSSI program, BAND collaboration), and 
by the Polish National Science Centre (NCN) under Contract No 
2018/31/B/ST2/02220.

\appendix

\section{Experimental landscapes of $\tilde{\delta}_{2\tau}$}
\label{sec:d2tau}

\begin{figure*}[htb]
    \centering{\includegraphics[width=0.8\textwidth]{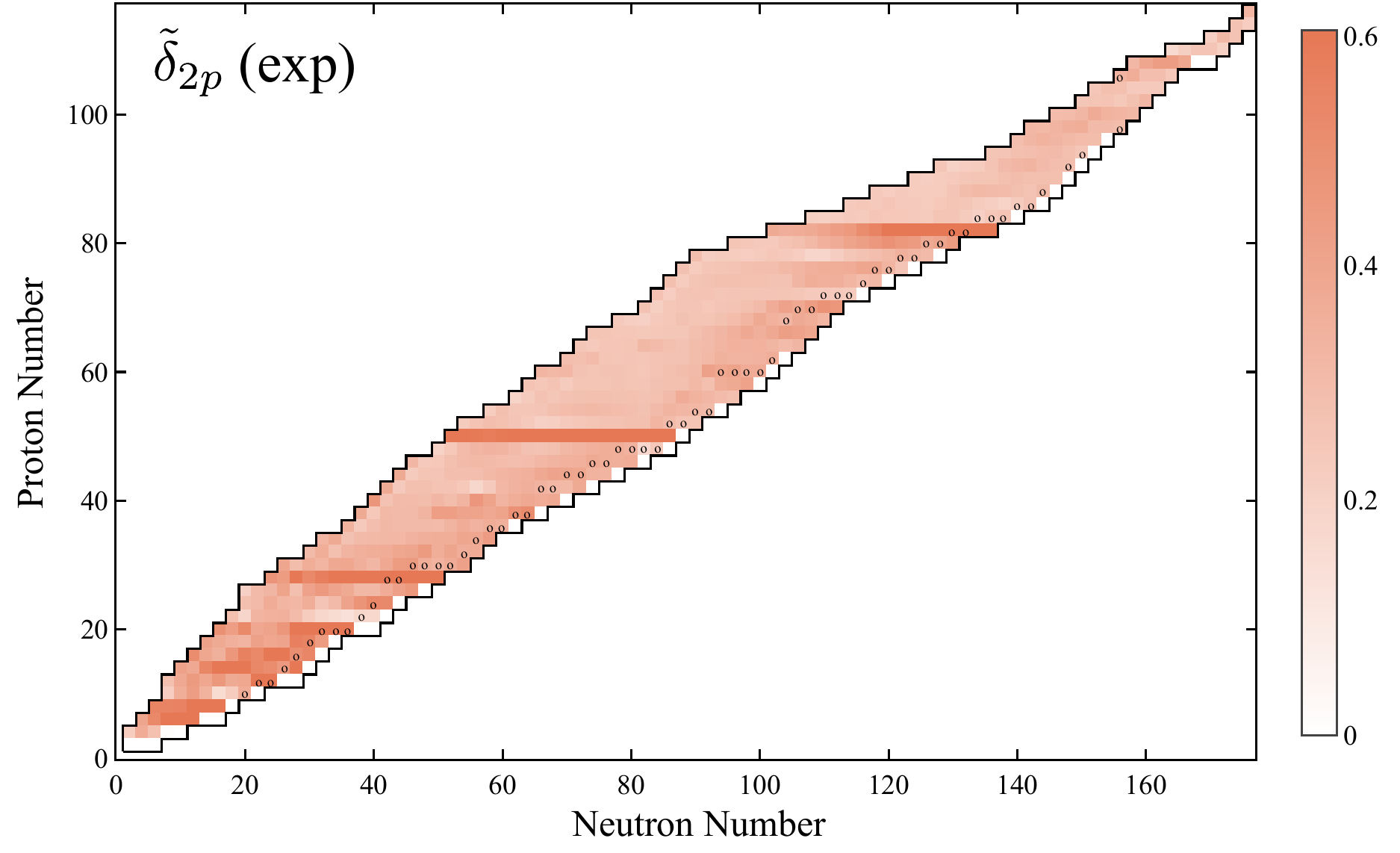}}
    \caption{Similar as in Fig. 2 but for experimental values of 
   the dimmensionless single-particle splitting $\tilde{\delta}_{2p}$. }
    \label{fig:SPESd}
\end{figure*}

\begin{figure*}[htb]
    \centering{\includegraphics[width=0.8\textwidth]{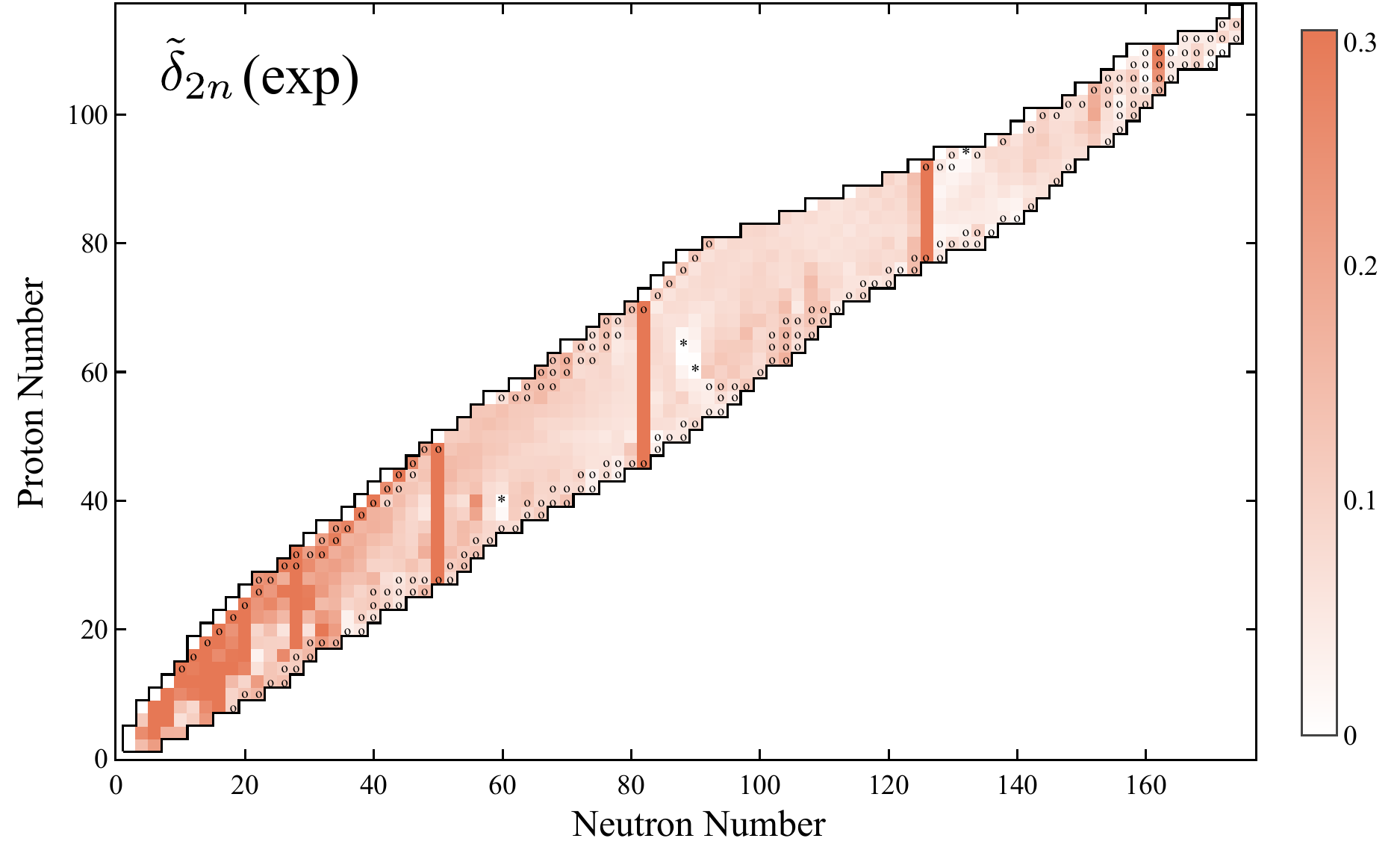}}
    \caption{Similar as in Fig. 3 but for experimental values of  the dimmensionless single-particle splitting   $\tilde{\delta}_{2n}$.}
    \label{fig:SNESd}
\end{figure*}

The two-nucleon shell gap indicators are usually defined as $\delta_{2n}=S_{2n}(N,Z) - S_{2n}(N+2,Z)$
and $\delta_{2p}=S_{2n}(N,Z) - S_{2p}(N,Z+2)$.
Figures \ref{fig:SPESd} and
\ref{fig:SNESd} show the dimensionless single-particle splitting 
$\tilde{\delta}_{2\tau}$
extracted from experimental values of $\delta_{2\tau}$ scaled as
 $\tilde{\delta}_{2\tau} \equiv \delta_{2\tau}/2\hbar\omega_{0}$,
where $\hbar\omega_{0}$ is given by Eq.~(\ref{eq:hw0}).

\section{The Bayesian Mass Explorer}
\label{sec:bmex}

\begin{figure*}[htb]
    \centering{\includegraphics[width=1.0\textwidth]{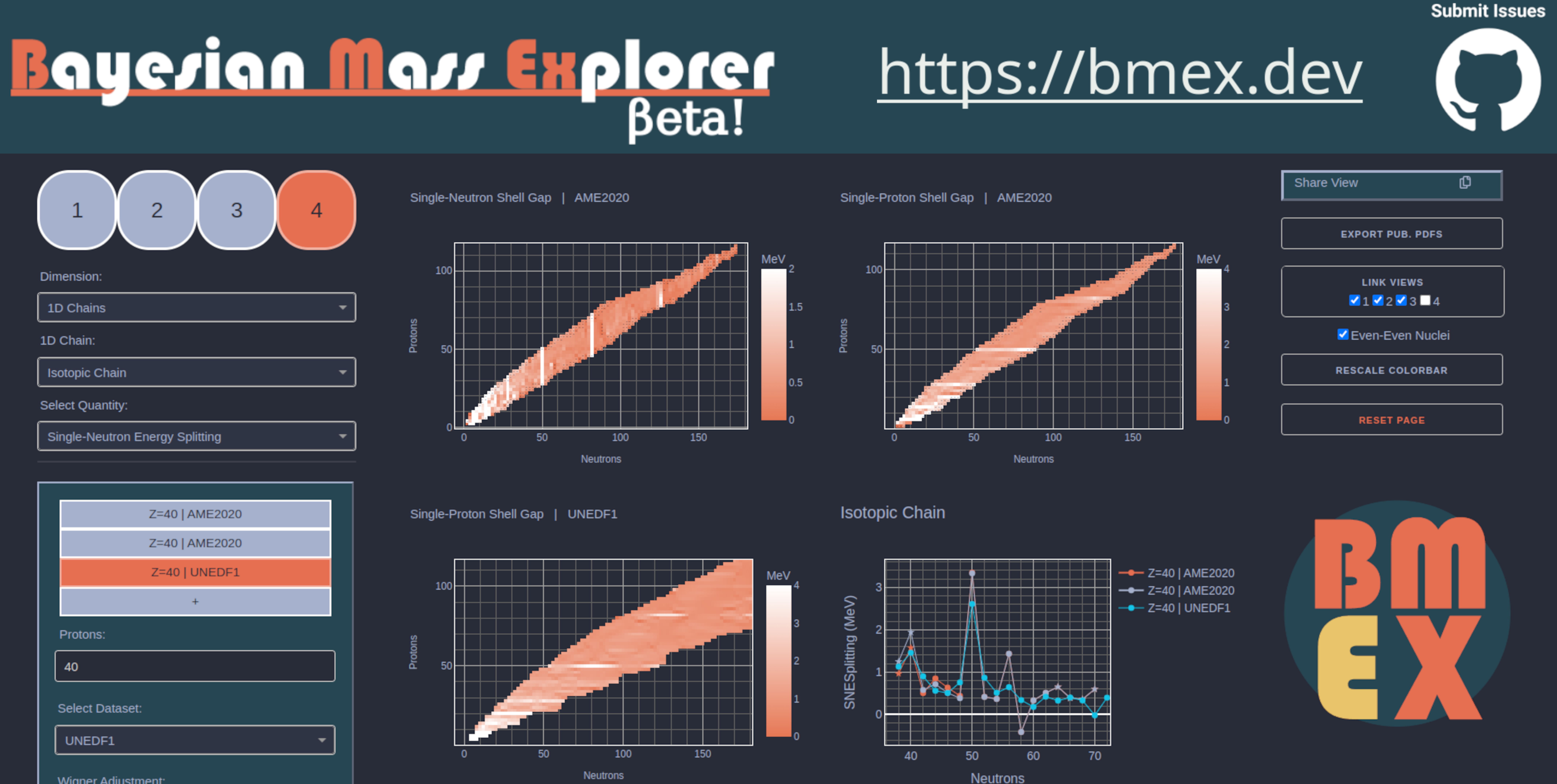}}
    \caption{Sample screenshot of the Bayesian Mass Explorer web interface in a configuration similar to what was used for the present investigation. The configuration and quantity of interest for each plot is accessed through the tab functionality on the left and additional data series can be added to the isotopic chains in the sidebar. The current view can be saved and shared using the ``Share View'' button on the right sidebar. Vector graphics in the  PDF format can also be exported through the web interface for easy integration into documents.}
    \label{fig:bmex}
\end{figure*}

The Bayesian Mass Explorer (BMEX) project aims to provide a user-friendly interface to theoretical model predictions with quantified uncertainties.
To enable this vision, BMEX utilizes a cloud-based infrastructure that allows for efficient data retrieval, plotting, and light computation to be performed server-side and then delivered to the user in their browser.
The web application uses the Plotly Dash framework and a Python backend that takes advantage of multiple server workers to better handle several users simultaneously.
The application is continuously built as a Docker container which eases deployment to arbitrary server architectures and can also easily be self-hosted locally for development or for local deployments.
Should the load on the server pass a certain threshold, independent instances of the container can also be deployed onto new servers and access to each one can be load balanced.
This improves availability and stability of the application, at the cost of needing to have a separately hosted database instance to manage user-saved sessions. 
A separate mechanism for saving user sessions is also implemented via link encoding, though this is less scalable and presents issues for backwards compatibility.
Figure\,\ref{fig:bmex} presents a sample screenshot of the BMEX web interface in a configuration similar to what was used for the present investigation of single-particle energy splittings.

\clearpage
%

\end{document}